\RequirePackage{fix-cm}
\documentclass[twocolumn]{svjour3}       
\smartqed  
\usepackage{graphicx}
\usepackage{array}
\usepackage{multirow}
\begin{document}

\title{A multiagent based framework secured with layered SVM-based IDS for remote healthcare systems
}
\subtitle{}

\titlerunning{A multiagent based framework secured with layered SVM-based IDS for remote healthcare systems}        

\author{MohammadReza Begli         \and
        Farnaz Derakhshan  
}


\institute{MohammadReza Begli \at
              MSc Graduated Student, Faculty of Electrical and Computer Eng., University of Tabriz, Tabriz, Iran \\
              \email{mohammadreza.begli95@ms.tabrizu.ac.ir}           
           \and
           Farnaz Derakhshan \at
              Assistant Professor, Faculty of Electrical and Computer Eng., University of Tabriz, Tabriz, Iran
              \email{derakhshan@tabrizu.ac.ir}
}

\date{Received: date / Accepted: date}

\maketitle

\begin{abstract}
Since the number of elderly and patients who are in hospitals and healthcare centers are growing, providing efficient remote healthcare services seems very important. Currently, most such systems benefit from the distribution and autonomy features of multiagent systems and the structure of wireless sensor networks. On the one hand, securing the data of remote healthcare systems is one of the most significant concerns; particularly recent types of research about the security of remote healthcare systems keep them secure from eavesdropping and data modification. On the other hand, existing remote healthcare systems are still vulnerable against other common attacks of healthcare networks such as Denial of Service (DoS) and User to Root (U2R) attacks, because they are managed remotely and based on the Internet. Therefore, in this paper, we propose a secure framework for remote healthcare systems that consists of two phases. First, we design a healthcare system base on multiagent technology to collect data from a sensor network. Then, in the second phase, a layered architecture of intrusion detection systems that uses Support Vector Machine to learn the behavior of network traffic is applied. Based on our framework, we implement a secure remote healthcare system and evaluate this system against the frequent attacks of healthcare networks such as Smurf, Buffer overflow, Neptune, and Pod attacks. In the end, evaluation parameters of the layered architecture of intrusion detection systems prove the efficiency and correctness of our proposed framework.
\end{abstract}

\section{Introduction}
\label{intro}
By growing the number of elderly and patients, providing healthcare services becomes difficult. Accordingly, using an intelligent system can facilitate these services. However, such a system should provide healthcare services without interrupting [1]. Consequently, appropriate hardware and software platform must be applied to design and implement healthcare systems. To this end, wireless sensors as a hardware platform can collect data of elderly or patients continuously and the system can provide the right response when it is necessary. On the one hand, the diversity of the data obtained from sensors in a healthcare system makes this data incorrect, unclear, and sometimes contradictory. On the other hand, sensors in an environment and patients cannot aggregate and make decisions based on the received data. To overcome this problem, using multiagent systems would be a good software model to manage these data [1].\\
\hphantom{th}Another main issue of healthcare systems is to secure personal data in wireless sensor networks of healthcare systems against attacks. Since healthcare systems are remotely manageable and based on Internet, they are vulnerable to frequent attacks of healthcare systems such as Denial of service (DoS) and User to Root (U2R) attacks. Consequently, attackers can make healthcare networks unavailable by DoS attacks, or a user who has access to data of a healthcare system can get a higher access level by committing User to Root attacks.
Therefore, in this research, we propose a framework that contains two steps. In the first step, we design a layered multiagent-based architecture. In the second step, we introduce a layered architecture to consider intrusion detection systems (IDS) for our agent-based healthcare system. Consequently, we classify agents according to their energy consumption and security of their data. Then, we allocate an intrusion detection system (IDS) [2] to each group of agents.\\
\hphantom{th}In multiagent systems (MAS), agents interact with each other autonomously. They do their jobs based on data gathered from sensors without interrupting other processes [1]. Consequently, using multiagent systems is a proper method to design our healthcare system.\\
\hphantom{th}In the second step of our framework, we assign intrusion detection systems to the agent-based healthcare system. Consequently, we introduce intrusion detection systems in the following paragraph briefly. Intrusion detection systems monitor and analyze activities in computer systems or computer networks to find out suspicious events that intrude them. Different kinds of Intrusion detection systems can be categorized base on the type of events that they monitor and their implementation [2]. In this paper, we use anomaly and misuse (signature-based) host-based Intrusion detection systems.\\
\hphantom{th}Therefore, the purpose of this paper is to propose a framework based on multiagent technology for collecting data from an environment and patients using a secured wireless sensor network at a hospital or a health facility. In this paper, we design an agent-based healthcare system to collect sensor networks data, provide healthcare services, and consider the security of healthcare sensor networks to confront possible attacks.\\
\hphantom{th}After this introduction, the rest of this paper is as follows: In the section 2, we shortly discuss related works, which are related to the background of our framework. In the section 3, we explain our proposed framework. The simulation results, which show and evaluate the performance of the framework, are presented in the section 4. In the section 5, first, we discuss our contribution to the improvement of healthcare systems, and then we compare our framework with some similar works in the past. Finally, in the section 6, we end this paper with a conclusion and mention some future works.

\section{Related works}
\label{sec:2}
This section presents some of the related works about remote healthcare systems, secured healthcare systems, and SVM-based intrusion detection systems. The basic concepts of this section are the prerequisite of our framework.
\subsection{Remote healthcare systems}
\label{sec:2-1}
Remote healthcare systems, which monitor the status of patients remotely and widely, consist of three main parts: 1) Some wireless sensors, 2) Sensor analysis and signal analysis to detect problems, and 3) Alert to healthcare staff. Aingeru [3, 4, 5], PANGEA [1, 6], MADIP [7], JTH [8], GerAmi [9], Koutkias [10], Kaluza [11], and Cervantes [12, 13] are examples of this kind of healthcare systems. Among these healthcare systems, PANGEA and MADIP systems are more recent than the others, and Aingeru system is similar to our proposed framework. Consequently, we introduce Aingeru, PANGEA, and MADIP in the following paragraphs bri-efly.\\
\hphantom{th}Aingeru healthcare system is a Java-based system. It also uses the Jade tool, which is a Java-based tool, as an agent platform. In this system, sensors are connected to PDA devices by Bluetooth 1.1 wireless connection, and PDA devices are connected to the Control Center by a GPRS connection. This healthcare system gives intelligent, comprehensive, and constant monitoring of the elderly and patients. Every cared person has a sensor-connected PDA, which is responsible for measuring vital parameters (such as heart rate, amount of oxygen in the blood). The system evaluates data locally (on a PDA) and delivers results to a relevant specialist in an emergency. The control center of a hospital collects all data from patient-connected sensors. Then it stores them in the central database. Physicians can get access to all relevant data of patients through a web-based program. The main components of this system are 1) PDA of a user that is for supervising users, 2) The control center that provides remote services, 3) The health center that gives medical services, and 4) Technical center, which is responsible for maintenance of all infrastructures.\\
\hphantom{th}The PANGEA multiagent healthcare system launched and tested at El Residencial La Vega healthcare facility in Salamanca, Spain for 8 months in 2014. In this system, relevant agents respond to condition of an environment based on data of sensors automatically. The following paragraphs are a brief description of the agents in the PANGEA system.\\
\hphantom{th}\textbf{Information Provider Organization:} This agent is the interface of the system to communicate with users.\\
\hphantom{th}\textbf{Home Automation Organization:} This part includes all agents that are responsible for controlling sensors in the environment. These sensors are a smoke detector, a temperature sensor, and a presence detection sensor. These agents send all data gathered from these sensors to the ZigBee Supervisor.\\
\hphantom{th}\textbf{ZigBee Supervisor Agent:} This agent is responsible for sending data to the database of the system.\\
\hphantom{th}\textbf{Information Agent:} This agent is the interface of the database.\\
\hphantom{th}\textbf{Locating Organization:}Incorporates all agents that are responsible for controlling sensors of patients to locate them.\\
\hphantom{th}\textbf{Caregiver Organization:} Every nurse and physician has one agent that aggregates data related to their activities.\\
\hphantom{th}The MADIP is a healthcare system in heterogeneous and wide-range networks such as the Internet or metrop-olitan and national networks. This system consists of two types of agents including static agents and mobile agents. Static agents offer resources and facilities for mobile agents. Mobile agents operate autonomously and communicate with static agents and host systems by using network infrastructure. The main parts of the MADIP system are as follows: User agent, Diagnostic agent, and resource agent as static agents, physician agent as a mobile agent, knowledge-based data server, and external services.\\
\subsection{Security of agent-based healthcare systems}
\label{sec:2-2}
In the following, we study earlier works about the security of agent-based healthcare systems. The priority concern of healthcare systems is sensitive data protection from unauthorized users. The simple approach in [14] uses the Access Control (AC) mechanism to address this problem. However, traditional AC models are inflexible and difficult to apply in healthcare systems, so this paper proposes two data filtering layers (policies) before returning results to the user, including access policy (public cloud) and privacy policy (private cloud). It introduces an access control model focusing on privacy protection for the healthcare systems. These two sets of policies protect patient records of healthcare systems, so they are accessible securely.\\
\hphantom{th}Fog computing provides storage, computing, and networking between end devices and the cloud [15]. Portable devices can be fog servers, where they can perform many tasks such as processing, visualization, and transferring data to the cloud. The main idea of Fog computing is to migrate some tasks of cloud data center tasks to the fog servers on the network edge. Also, Fog computing has proven to be extremely useful in time-sensitive applications such as health care applications [16–18]. Consequently, the use of Fog computing improves security because of its proximity to the devices [19] and accelerates real-time data processing of applications [19]. Since medical information is considered sensitive data, the privacy issue in e-health systems is vital. This issue has been taken care of in [20] by prohibiting unauthorized access and encrypting information before sending it to storage.\\
\hphantom{th}The biometric features acquired from biosignals are distinctive and random. Consequently, they can develop security mechanisms for smart healthcare systems. In [21] the Inter-Pulse Intervals (IPIs) extracted from heartbeats are used to generate secret keys to secure WBSNs. These keys are distributed to nodes within and outside the WBSNs to secure data transmission in remote healthcare platforms.\\
\hphantom{th}In [22] a three-factor authentication protocol, which is based on Shamir’s threshold scheme, increases the security and privacy of data in healthcare systems and minimizes the communication cost. \\
\hphantom{th}The architecture of the healthcare system proposed in [23] is composed of three layers. The first layer, which is the lowest layer and called Sensor Network Tier layer, includes two types of sensors. The first type is the wearable sensor system that pairs with biological sensors (belt attached to caregivers). The second type is the sensors that are located in a building and are responsible for transmitting data of an environment using a wireless or wired network with the ZigBee protocol [24]. The second layer, called mobile computing network tier, includes mobile devices such as PDA devices and laptops that connect to a fixed station using a network infrastructure or a multi-hub network. The third layer, back-end network tier, includes stationary stations and servers that give application layer services for lower layers and is responsible for processing and saving the data received from mobile devices.\\
\hphantom{th}The purpose of this architecture is to keep the confidentiality of the data stored in each layer using a digital signature or proper encryption. For example, in the network layer of this sensor network, AES encryption secures the communication between each pair of nodes. Access of devices in the second layer to the data of the first layer is limited, so access is available after authentication. Furthermore, each device in the second layer has a public key and private key pair that is assigned by a third-party entity from the third layer (local server). Hosts in the second layer use this key pair to encrypt access to the first layer.\\
\hphantom{th}In [25], the approach is to balance the performance and the security of a sensor-based healthcare system. In this way, this sensor network consists of three parts as follows: the Medical Sensor Network (MSN), the Patient Area Network (PAN), and a service center. Also proposed security architecture consists of three separate and complementary layers, so there is a security structure for each part of the network.\\
\hphantom{th}The MSN layer includes environment sensors such as temperature and humidity sensors and mobile devices such as PDA devices. Symmetric encryption secures the communication between each pair of nodes in this layer.\\
The security layer of PAN, which includes patient-conne-cted sensors, secures medical data associated with MSN-layer nodes (such as physicians and nurses) and a service center. This security layer authenticates devices connected to patients by using a private key.
The corresponding security layer of the service center consists of a public key and a certificate. Therefore, components of the PAN and the MSN can receive a pair of private and public keys and a unique identification number (ID) by having the public key of the service center certificate.\\
\hphantom{th}To conclude, encryption algorithms are proposed to secure remote healthcare systems in these pieces of research. However, the cryptographic algorithms cannot confront attacks except eavesdropping and data modification, which means these systems are vulnerable to other common attacks of remote healthcare systems such as buffer overflow. Therefore, in this research, we intend to improve this security weakness of earlier remote healthcare systems by considering the energy consumption of sensors and the sensitivity of their data.\\
\subsection{SVM-based IDS}
\label{sec:2-3}
In our proposed framework, we intend to improve the security weakness of remote healthcare systems mentioned in the previous paragraph. For this purpose, we use intrusion detection systems in our proposed framework. We explain the mechanism of intrusion detection systems in the next section. These intrusion detection systems use the support vector machine to classify network traffic. Therefore, in this section, we introduce earlier works about intrusion detection systems to prove the advantages of SVM in comparison to other network traffic classifiers.\\
\hphantom{th}In [26], machine-learning methods such as Optimal Power Flow (OPF), SVM, and K-Nearest Neighbors (KNN) collaborate to find abnormal traffic in computer networks. The network traffic contains information such as source and destination IP address, packet count, service type, protocol type, source port, destination port, start time, and end time of a session between two hosts in a network. The focus of this research is on DoS, Brute Force, and Port Scanning attacks. The IDS in this research detects all these three types.\\
\hphantom{th}In [27], the IDS consists of the PCA (principal components analysis) method and the SVM. The approach of this research is to select a suitable set of features. NSL-KDD standard data set is used to evaluate the performance of the proposed IDS. \\
\hphantom{th}Research [28] describes SVM as one of the best methods to detect abnormal behaviors in networks and proposes an SVM based IDS. It mentions that SVM has a high computational cost, although if the number of features of the training set decreases, the SVM machine will perform better. \\
\hphantom{th}Research [29] proposes an SVM-based IDS to identify attacks in enterprise networks. The features are selected by information techniques to learn multiple SVM. According to [29], this method has a higher detection rate than the artificial neural network. Furthermore, some other works in this field prove that SVM performs better than different network traffic classifiers [30, 31].\\
\hphantom{th}According to the earlier research, which we reviewed in this section, we got enough background knowledge for our proposed framework. Therefore, in the next section, we describe our framework.\\
\section{System overview}
\label{sec:3}
In this section, we introduce our framework for multia-gent-based healthcare systems that facilitates the provision of healthcare services and secures the wireless sensor network of healthcare systems against attacks simultaneously.\\
\hphantom{th}Multiagent systems help us to integrate data in remote healthcare systems due to the distribution of the data. Each agent has a specific responsibility and interacts with other agents to manage and analyze the data to improve services. Therefore, we design our architecture base on multiagent systems. Consequently, as shown in Figure 1, the first step is to define and plan agents of our system. The second step is to classify agents base on their data and energy capacity. Then, for each group of agents, we consider a proper IDS to secure the sensor network of our healthcare system against attacks.\\
\hphantom{th}Therefore, in the first part of this section, we design our multiagent framework and plan agents, capabilities, and goals of them. In the second part of this section, we classify agents according to their data and energy capacity. Then we add a suitable IDS to each group of agents.\\
\begin{figure}
  \includegraphics{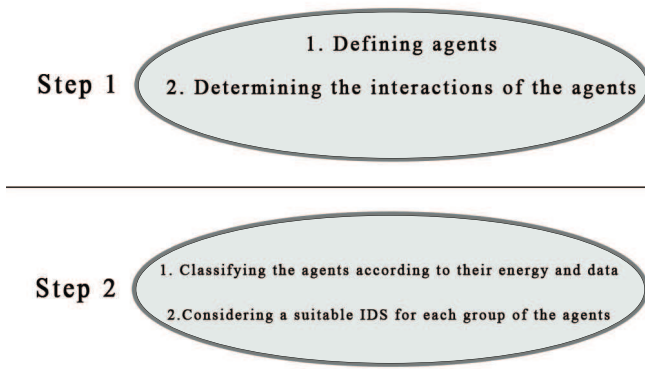}
\caption{The guideline of our proposed framework}
\label{fig:1}       
\end{figure}
\subsection{Multiagent design}
\label{sec:3-1}
In this section, we present the multiagent design of our healthcare system. Figure 2, which represents all agents and interactions between them in Tropos methodology, shows the multiagent architecture of our proposed framework. Tropos [32–34] is a software engineering method for designing multiagent systems. This methodology covers all concepts of agents such as goals, programs, and tasks at the development stage. It also focuses on software requirements. The Tropos methodology is about agents and all its related ideas and analysis of primary software needs. The actor diagram is for the stage of determining the basic requirements of the software. In the Tropos methodology, a circle expresses an agent, an ellipsoid shows a goal, a cloud shows a soft goal, and a dashed line implies that there are several sub-systems of the same kind. A soft goal is an ambiguous goal in a system that there is no clear benchmark for achieving. Actors in multiagent systems are agents that play a role in a system. Therefore, according to Figure 2, there are five actors in our proposed framework. In the following paragraphs, we introduce these agents (actors).\\
\hphantom{th}\textbf{Patient agent} manages all data related to patients. These data include the profile of patients, heart rate, blood pressure, body temperature, and the location of patients. The database center saves these data at the start of admission to a health facility center. Also, real-time sensors of patients collect other necessary data such as blood pressure, heart rate, and body temperature. Presence sensors attached to patients also find the location of patients and detect the presence of patients in their beds. This agent detects the condition of patients after the integration of health monitoring data. After reviewing the collected data, if the patient agent diagnoses the health condition of a patient as unsuitable, it will alert the nurse agent. Also, the patient agent can send his/her request to the nurse agent.\\
\hphantom{th}\textbf{Nurse agent} manages activities of nurses according to the priorities of them. This agent reports the status of patients to a specific nurse and warns in an emergency. It also requests nurses to do activities and collects reports from them.\\
\hphantom{th}\textbf{Physician agent} Physicians do their jobs with the help of this agent on portable devices such as tablets, laptops, and smartphones. This agent helps them to monitor the status of patients and receive reports of nurses. Physicians check the condition of patients and the activities of nurses with the help of the physician agent virtually. They can also check the health condition of patients and send out all activities that nurses should do for patients with the help of this agent.\\
\hphantom{th}\textbf{Ambient agent} controls sensors in an environment include smoke detection sensors and temperature sensors. This agent identifies the condition of an environment after collecting the data of these sensors. If this agent detects an emergency in an environment such as a fire near to patients, it will alert nurses.\\
\hphantom{th}\textbf{Database agent} is the database interface. It maintains and manages the profile of patients. This agent keeps alert messages with the feedbacks of nurses.\\
\begin{figure*}
  \includegraphics[width=1\textwidth]{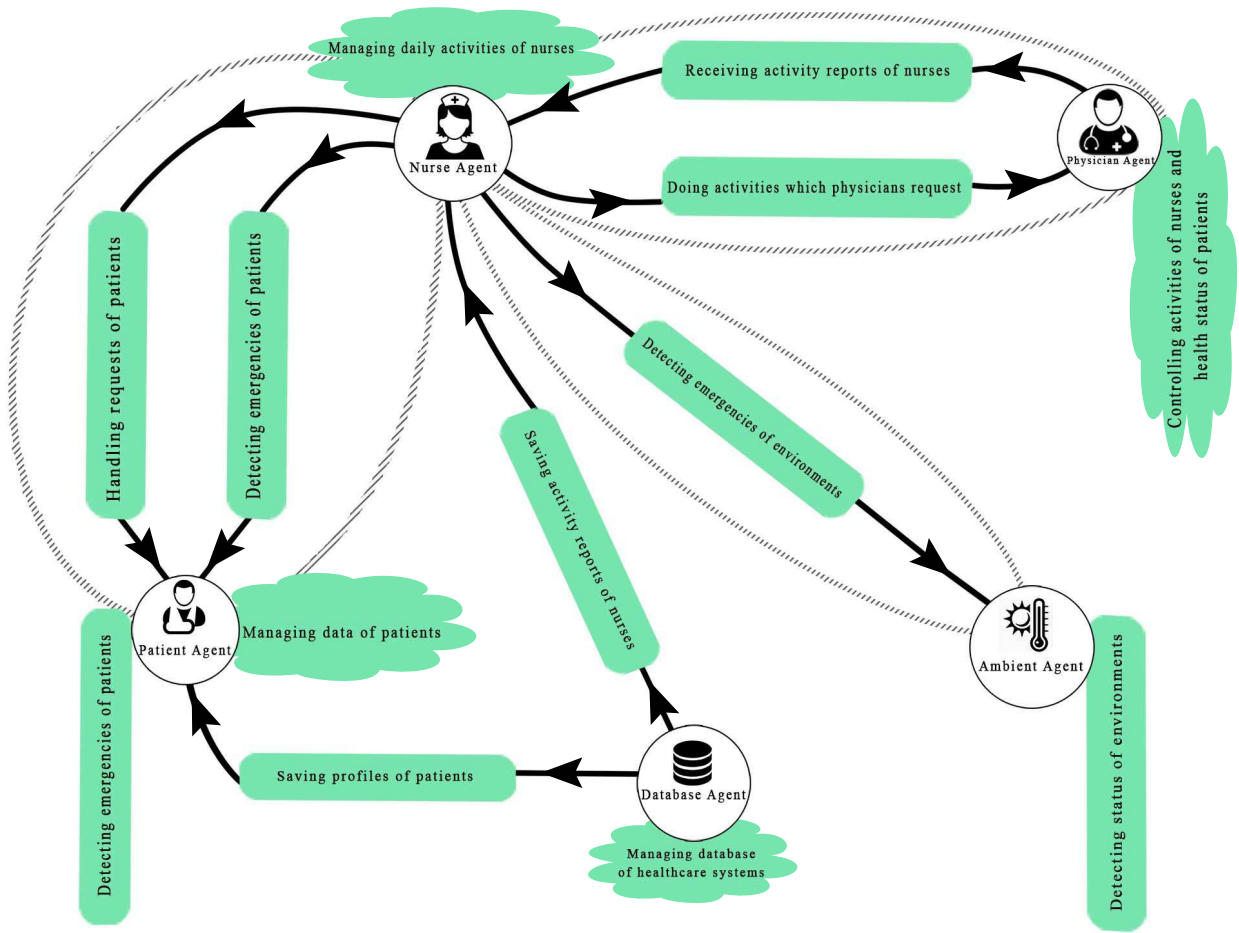}
\caption{The actor diagram of the proposed framework in Tropos methodology}
\label{fig:2}       
\end{figure*}
\hphantom{th}According to Figure 3, after the admission of patients to a healthcare center, the interactions between the agents begin as follows:\\
\hphantom{th}1. After collecting data from patients, the patient agent sends identity information of patients, such as name, age, and date of entrance, to the database agent. The database agent saves them in the database.\\
\hphantom{th}2. The patient agent analyzes data, including blood pressure, heart rate, and body temperature, to find the health condition of patients. After analyzing the obtained data, if the patient agent detects that the health condition of a patient is unsuitable, it will alert the nurse agent.\\
\hphantom{th}3. The patient agent sends the request of the patient to the nurse agent.\\
\hphantom{th}4. The ambient agent collects the data of ambient sensors, including smoke detector sensors and temperature sensors, to detect the status of an environment. In an emergency, this agent alerts the nurse agent.\\
\hphantom{th}5. The nurse agent sends the activity reports of nurses to the physician agent after nurses react to received alerts.\\
\hphantom{th}6. The nurse agent sends activity reports to the database agent to save the alerts and their reactions.\\
\hphantom{th}7. The physician agent can request the nurse agent to do treating activities.\\
\hphantom{th}8. The physician agent can ask the patient agent to send the health condition of patients.\\
\hphantom{th}9.	The patient agent responds to the request of the physician agent by sending blood pressure, heart rate, body temperature, etc. \\
\begin{figure*}
  \includegraphics[width=1\textwidth]{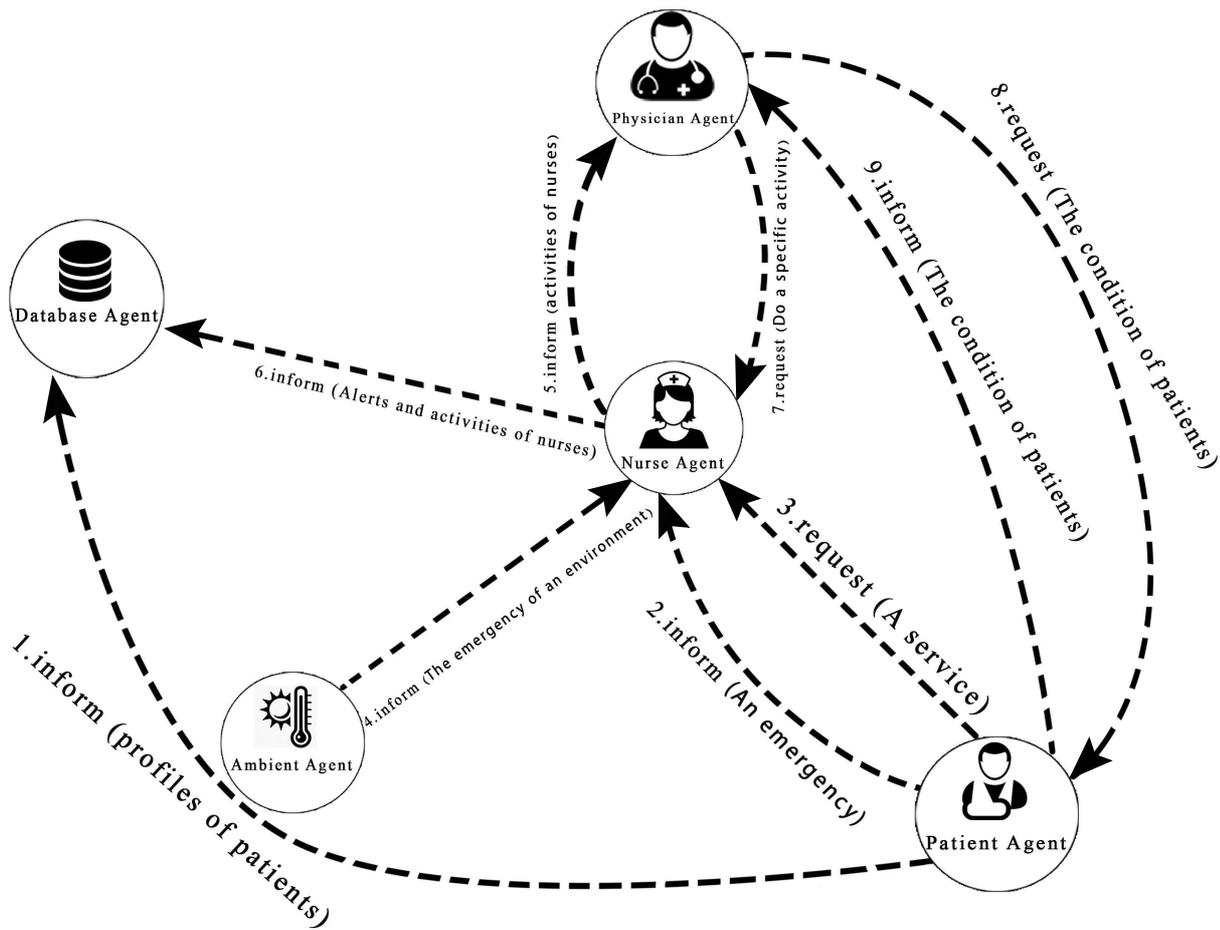}
\caption{The association of the agents in the proposed framework}
\label{fig:3}       
\end{figure*}
\subsection{Layered architecture of intrusion detection systems}
\label{sec:3-2}
As we mentioned before, the second step is securing the healthcare system. We use a clustered wireless sensor network [35] (one of the most common network topologies in sensor networks) in our healthcare system. Because of the difference in activities of each sensor, the probability of an attack is different for each one. Therefore, in the following paragraph, we discuss the mechanism of the IDS for clustered wireless sensor networks in the research [36].\\
\hphantom{th}The capabilities of sensors in clustered sensor networks are heterogeneous [37, 38]. Sinks need to collect data from all clusters in these networks, so they have more processing power and use more energy than sensors. Cluster sensors also have more processes than ordinary sensors, so they use more energy. Other sensors do not have the same capacity as cluster heads because they only collect ambient data within their scope. Consequently, the capabilities of sinks are more than cluster heads, and the capabilities of cluster heads are more than other sensors. As a result, sinks and cluster heads are more affected by attacks than ordinary sensors, and they should have better security measures.\\
\hphantom{th}We also arrange the agents of our framework and use an IDS suitable for each group of agents with similar reasons. As shown in Figure 4, first, we classify the agents based on the sensitivity of their data and the energy level of their corresponding sensors. In this case, the database agent and other agents (that have data from aggregated sensors) have better security measures than those with restricted data. \\
\hphantom{th}The patient agent collects the health data of patients, and the ambient agent collects the data of ambient sensors. Therefore, these two agents have restricted data, which are not valuable for attackers. Consequently, they rarely attack these agents. Furthermore, the wearable sensors of patients have limited energy, so we consider an anomaly IDS for these agents. Anomaly intrusion detection systems, which have a high detection rate, create a model of normal traffic, and detect network attacks by comparing network traffic with the model. However, anomaly intrusion detection systems identify some normal traffic of networks which has different patterns from the rest as an attack wrongly, so they have a high false-positive rate and low accuracy [36]. Also, they have a low computational cost and low power consumption because of their simple structure, which makes them suitable for both the patient and the ambient agents.\\
\hphantom{th}The nurse agent collects the health status of patients and the status of the ambient agents. Then, in an emergency, the nurse agent alerts nurses, and after recording feedbacks of nurses, the nurse agent sends a record of alerts to the database agent. Besides, the physician agent can ask the nurse agent to do a specific action, such as sending the health status of patients. Consequently, the nurse agent and the physician agent collect data from patients and environments, which makes them more likely to be attacked than the patient agent and the ambient agent. The physician agent and nurse agent do not collect data directly from sensors, and they run on mobile devices such as mobile phones, so they have more energy supply than the ambient agent and the patient agent. As a result, we consider a misuse intrusion detection system for the physician agent and the nurse agent. Misuse intrusion detection systems detect the type of attacks in addition to the normal, and attack traffic, so they have a more complex structure, higher processing cost, and more energy consumption than anomaly intrusion detection systems.\\
\hphantom{th}The database agent saves profiles of patients, alerts of the patient agent and the ambient agent to the nurse agent, and reports of nurses' reactions. This agent collects and saves all data of sensors in our healthcare system. Consequently, the database agent is more likely to be attacked than the rest of the agents and requires a better intrusion detection system than the two lower layers in Figure 4. As a result, we consider a hybrid intrusion detection system for this agent. In hybrid intrusion detection systems, an anomaly intrusion detection system and a misuse intrusion detection system complement each other. The anomaly intrusion detection system, which has a high detection rate, detects anomaly traffic. Then the misuse intrusion detection system corrects traffic, in which the anomaly intrusion detection system detects abnormal wrongly due to its low accuracy [39]. Finally, the decision-making unit determines the type of traffic. Therefore, hybrid intrusion detection systems use the high detection rate of anomaly intrusion detection systems and the high accuracy of misuse intrusion detection systems.\\
\begin{figure}
  \includegraphics{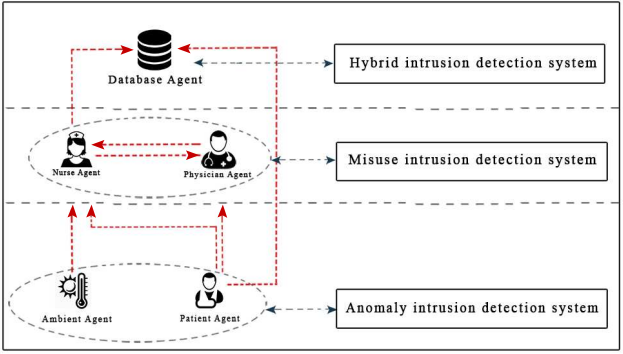}
\caption{Layered architecture of the intrusion detection systems in the proposed framework}
\label{fig:4}       
\end{figure}
\hphantom{th}Table 1 shows the rules of the decision-making unit of the hybrid intrusion detection system. As we mentioned before, anomaly intrusion detection systems have a high detection rate in detecting normal traffic. In the first rule, this feature helps to distinguish normal traffic. We also mentioned that anomaly intrusion detection systems have low accuracy in detecting attacks. The high accuracy of misuse intrusion detection systems modifies this low accuracy in the second and third rules. Consequently, if the misuse IDS does not find an attack that the anomaly IDS detects, so this traffic is normal. If the misuse IDS detects the class of an attack that the anomaly IDS find, then it is an attack.
\begin{table}
\centering
\caption{Rules of the hybrid IDS in our framework}
\label{tab:1}
\resizebox{\linewidth}{!}{%
\begin{tabular}{>{\hspace{0pt}}p{1\linewidth}}
\hline
\textbf{Rules}                                                                                                                                                                              \\
\hline
If the anomaly IDS detects traffic as normal, traffic will be normal.                                                                                                                       \\
\hline
If the anomaly detection detects an attack and the misuse detection does not detect any attack, then it is not an attack and it is an incorrect classification.                             \\
\hline
If the anomaly detection detects an attack and similarly the misuse detection detects an attack, then it is an attack and the decision-making unit will determine the class of the attack.  \\
\hline
\end{tabular}
}
\end{table}
\hphantom{th} Hybrid intrusion detection systems have more computational cost and more energy usage than anomaly intrusion detection systems and misuse intrusion detection systems because they use two different types of intrusion detection systems at the same time. The database agent has an unlimited energy supply (the database server works with electrical power supply.), so hybrid intrusion detection systems fit the database agent perfectly. \\
\hphantom{th}So far, we have introduced details of our proposed framework. In the next section, we discuss how to implement the agents in this framework. Additionally, we evaluate the intrusion detection system in each layer of the framework to prove the correctness of the agents' classification and assigning an IDS to each layer.
\section{Experiment}
\label{sec:4}
After the phase of analysis and design, we implement and test our framework. In this section, we focus on the evaluation and implementation results of our healthcare system. Also, we examine our solution by using a dataset, and then we present the run-time and the memory consumption of the intrusion detection systems. Finally, we compare previous works in terms of the security of healthcare systems with our proposed method.
\subsection{Implementation of our multiagent-based healthcare system}
\label{sec:4-1}
We implement the agents of our proposed framework with JADE tool (version 4.4) in Java (JDK 8) by using NetBeans 8.2. First of all, we create the required agents from the classes which we have implemented with JADE in Java. As we mentioned previously, we use the following agents in this study: Database agent, Physician agent, Nurse agent, Ambient agent, and Patient agent. The database agent saves and manages the profile of patients under the care and keeps alert messages and feedbacks of nurses. In this study, we create a dataset consist of 1000 patients. These samples of patients include blood pressure, body temperature, heart rate, location of patients, ambient emergency, and ambient temperature samples. The patient agent and the ambient agent can access these data. According to the rules of the patient agent and ambient agent, if the health data of patients and the situation of an environment become abnormal, these agents will alert the nurse agent. We also implement other interactions between the agents, which we described previously.
\subsection{The data collection of the intrusion detection systems}
\label{sec:4-2}
As we discussed in Section 2, in earlier research, proper encryption algorithms are proposed in three layers to secure healthcare systems. However, these algorithms cannot confront common attacks of healthcare systems except eavesdropping and data modification. Therefore, our focus in this study is to propose a healthcare system framework that has tight security against common healthcare system attacks. Consequently, we first study the usual attacks on healthcare systems, which previous research mentions. Then we determine their types according to the general classification of network attacks to get an appropriate sample of network traffic to evaluate the intrusion detection systems of our framework.\\
\hphantom{th}Network attacks are classified into four groups [40, 41] as follows: Denial of Service (DoS), Remote to Local (R2L), User to Root (U2R), and Probe. Also, according to [42-44] spoofing attack, impersonation, the elevation of privileges, DoS attack, eavesdropping, and data modification are the usual attacks of healthcare networks. These attacks, except eavesdropping and data modification, are mainly U2R and DoS attacks.\\
\hphantom{th}To implement and evaluate the intrusion detection systems of our proposed framework, we use NSL-KDD [45] dataset. This dataset solves some of the inherent problems of the KDD'99 and is a newer version of that. Therefore, the NSL-KDD has some benefits over the KDD datasets as follows: 1) It does not include redundancy of the data or duplicate records in the train set, so the classifiers are not biased towards the more frequent data. 2) There are no duplicate records in the proposed test sets. Therefore, the performance of the learners is not biased by the methods that have better detection rates on frequent data. 3) The number of selected records from each difficulty group is proportional to the percentage of the related data in the original KDD data set. 4) The number of records in the train and test sets is reasonable, which makes it affordable to run experiments on this dataset.\\
\hphantom{th}To cover more attack types of healthcare systems, we extract U2R and DoS attacks from the attack samples of the NSL-KDD dataset by writing a query in Excel for each attack. Consequently, as Table 2 shows, we have ten datasets of the attacks, and a dataset of normal traffic from the NSL-KDD.\\
\hphantom{th}Then, we implement a program in Java environment to create our dataset consists of ten attacks. This dataset contains 10,000 samples consist of 4,000 samples of normal traffic and 300 samples from every ten attacks. We use three datasets because we use three different layers of the intrusion detection systems in our proposed framework.\\
\hphantom{th}Table 3 and Table 4 show the type of traffic samples and the number of each one in the test set and the train set for the anomaly, the misuse, and the hybrid intrusion detection systems, respectively. We obtain this information by analysis of the datasets with Weka tool (version 3.9) [46, 47]. In the next part of this section, we implement the intrusion detection systems by using SVM and evaluate them with the datasets.\\
\subsection{The implementation results of the intrusion detection systems}
\label{sec:4-3}
We implement the intrusion detection systems of our proposed framework with R tool (version 3.5.0) [48]. These intrusion detection systems classify the traffic of the network by using support vector machine (SVM) [30]. SVM is one of the most efficient machine learning algorithms in pattern recognition problems, such as classifying traffic of networks and image processing.\\
\hphantom{th}In the IDS of each layer of our framework, SVM detects the pattern of abnormal traffic from normal traffic, which is non-linear patterns. Non-linear patterns cannot be easily distinguished and separated, unlike linear patterns, which are easily distinguishable. So non-linear patterns need some process to become easily separable. Consequently, the Kernel function maps the original data to a higher dimensional space to classify non-linear patterns [30].\\
\hphantom{th}In the R tool, functions of pattern recognition problems are available in the caret package, and the kernel functions are in the kernlab package. Therefore, we use these two packages in the implementation of the intrusion detection systems of our framework. We use the ksvm function from the caret package to use SVM as a network traffic classifier. The first argument of ksvm function is the attribute of classification. In this study, it is traffic types that can be normal or the name of the attack (abnormal). The second argument declares the dataset that SVM classifies. The last argument names the type of the ksvm function. We use the c-svc type, which is for solving data classification problems. Finally, we name the kernel function. The kernel function, which we use to classify our network traffic, is the radial basis function (RBF in R). After we implement the intrusion detection systems of each layer, we evaluate the parameters of each layer's IDS. In the following, we discuss and conclude about these parameters.\\
\begin{table}
\centering
\caption{DoS and U2R attacks of the NSL-KDD dataset}
\label{tab:2}
\resizebox{\linewidth}{!}{%
\begin{tabular}{>{\hspace{0pt}}p{0.548\linewidth}>{\hspace{0pt}}p{0.452\linewidth}}
\hline
\textbf{ Attack name } & \textbf{ Attack type }  \\
\hline
~ ~ ~ ~ Back           & ~ ~ ~ DOS               \\
~Buffer overflow       & ~ ~ ~ U2R               \\
~ ~ ~ ~ land           & ~ ~ ~ DOS               \\
~ ~loadmodule          & ~ ~ ~ U2R               \\
~ ~ ~ neptune          & ~ ~ ~ DOS               \\
~ ~ ~ ~ ~perl          & ~ ~ ~ U2R               \\
~ ~ ~ ~ ~pod           & ~ ~ ~ DOS               \\
~ ~ ~ ~rootkit         & ~ ~ ~ U2R               \\
~ ~ ~ ~smurf           & ~ ~ ~ DOS               \\
~ ~ ~teardrop          & ~ ~ ~ DOS               \\
\hline
\end{tabular}
}
\end{table}
\begin{table}
\centering
\caption{The number of traffic samples from each type, in the dataset of the anomaly IDS}
\label{tab:3}
\resizebox{\linewidth}{!}{%
\begin{tabular}{>{\hspace{0pt}}p{0.408\linewidth}>{\hspace{0pt}}p{0.304\linewidth}>{\hspace{0pt}}p{0.281\linewidth}}
\hline
\textbf{Traffic type} & \multicolumn{2}{>{\hspace{0pt}}p{0.585\linewidth}}{\textbf{~ ~ ~ ~ ~ ~ ~Dataset }}  \\
\hline
                      & Train set & Test set                                                            \\
\hline
~ ~ Attack            & ~ ~1471   & ~ ~ 997                                                             \\
\hline
~ ~Normal             & ~ ~4000   & ~ ~4000                                                             \\
\hline
~ ~ ~Total            & ~ ~5471   & ~ ~4997                                                             \\
\hline
\end{tabular}
}
\end{table}
\begin{table}
\centering
\caption{The number of traffic samples from each type, in the dataset of the misuse IDS and hybrid IDS}
\label{tab:4}
\resizebox{\linewidth}{!}{%
\begin{tabular}{>{\hspace{0pt}}p{0.204\linewidth}>{\hspace{0pt}}p{0.373\linewidth}>{\hspace{0pt}}p{0.219\linewidth}>{\hspace{0pt}}p{0.204\linewidth}}
\hline
\multicolumn{2}{>{\hspace{0pt}}p{0.577\linewidth}}{\textbf{~ ~ ~ ~ ~ Traffic type }} & \multicolumn{2}{>{\hspace{0pt}}p{0.423\linewidth}}{\textbf{~ ~ ~ ~ ~Dataset }}  \\
\hline
\multirow{11}{0.204\linewidth}{\hspace{0pt}Attack} &                                 & Trainset & Testset                                                              \\
\cline{2-4}
                                                   & ~ ~ ~neptune                    & ~ ~300   & ~ 300                                                                \\
\cline{2-4}
                                                   & ~ ~ ~ ~back                     & ~ ~300   & ~ 300                                                                \\
\cline{2-4}
                                                   & ~ ~ ~ smurf                     & ~ ~300   & ~ 300                                                                \\
\cline{2-4}
                                                   & ~bufferoverflow                 & ~ ~ 30   & ~ ~20                                                                \\
\cline{2-4}
                                                   & ~ ~ ~ ~ pod                     & ~ ~201   & ~ ~41                                                                \\
\cline{2-4}
                                                   & ~ load module                   & ~ ~ ~9   & ~ ~ 2                                                                \\
\cline{2-4}
                                                   & ~ ~ ~ ~ perl                    & ~ ~ ~3   & ~ ~ 2                                                                \\
\cline{2-4}
                                                   & ~ ~ ~ ~ land                     & ~ ~ 18   & ~ ~ 7                                                                \\
\cline{2-4}
                                                   & ~ ~ ~ rootkit                    & ~ ~ 10   & ~ ~13                                                                \\
\cline{2-4}
                                                   & ~ ~ ~teardrop                    & ~ ~300   & ~ ~12                                                                \\
\hline
~Normal                                            & ~ ~ ~ Normal                      & ~ 4000   & ~4000                                                                \\
\hline
~ Total                                            &                                 & ~ ~5471  & ~4997                                                                \\
\hline
\end{tabular}
}
\end{table}
\hphantom{th}Table 5 shows the average runtime and memory consumption of the intrusion detection systems of the proposed framework after ten runs. The information in this table shows the correctness of considering the intrusion detection system for each layer of the proposed framework. As we mentioned earlier, the structure of the anomaly IDS is simpler than the misuse and the hybrid IDS. Therefore, it has less runtime and memory consumption than the misuse and the hybrid IDS in Table 5. It means it is suitable for the patient and the ambient agents, which have limited energy. Misuse (signature-based) IDS, which we consider for the nurse and the physician agents, has more runtime and memory consumption than the anomaly IDS. The hybrid IDS, which we consider for the database agent, uses two intrusion detection systems at the same time. Therefore, it has the most runtime and memory consumption.\\
\hphantom{th}Table 6 shows the detection rate and false-positive rate of the intrusion detection systems of the proposed framework. The information in this table shows the correctness of considering the intrusion detection system for each layer of the proposed framework.\\
\hphantom{th}According to table 6, the anomaly IDS has a high detection rate and a high false-positive rate (low accuracy) than the misuse IDS. It means the anomaly IDS is suitable for the patient and the ambient agents because they have restricted data of the network, so attackers rarely attack these agents. The misuse IDS has a lower false-positive rate (higher accuracy) than the anomaly IDS. It means it is suitable for the nurse agent and the physician agent, which have more data of the network than the patient agent and the ambient agent, so they are more likely to be attacked. In the hybrid IDS, the anomaly IDS and the misuse IDS complement each other. In other words, as table 6 shows, it benefits from the high detection rate of the anomaly IDS and the high accuracy of the misuse IDS. Consequently, it is suitable for the database agent, which is highly potential to be attacked.\\
\begin{table}
\centering
\caption{Average runtime and memory usage of the intrusion detection systems used in the framework}
\label{tab:5}
\resizebox{\linewidth}{!}{%
\begin{tabular}{>{\hspace{0pt}}p{0.213\linewidth}>{\hspace{0pt}}p{0.292\linewidth}>{\hspace{0pt}}p{0.494\linewidth}}
\hline
\textbf{~ ~IDS} & \textbf{Runtime (s)} & \textbf{Memory usage (MB)}  \\
\hline
Anomaly         & ~ ~ ~3.001           & ~ ~ ~ ~ ~275.088            \\
\hline
~Misuse         & ~ ~ ~7.784           & ~ ~ ~ ~ ~285.486            \\
\hline
~Hybrid         & ~ ~ 11.474           & ~ ~ ~ ~ ~286.138            \\
\hline
\end{tabular}
}
\end{table}
\section{Our Contributions}
\label{sec:5}
As we mentioned in Section 2, spoofing, impersonation, the elevation of privileges, DoS attacks, eavesdropping, and data modification are the usual attacks of healthcare networks. According to some earlier research that we discussed before, proper encryption algorithms were proposed in three layers to secure healthcare systems. These algorithms cannot confront attacks except eavesdropping and data modification. Therefore, the main advantages of our proposed framework in comparison with earlier works are as follows: 1) Designing a healthcare system by using the multiagent technology to collect data of the healthcare network. It means we have one agent for each patient and an environment with specific rules. 2) Considering security measures to confront possible attacks other than eavesdropping, besides the provision of healthcare services. 3) Allocation of SVM-based intrusion detection systems appropriate to each group of the agents to save energy and computational cost of the network.
\begin{table}
\centering
\caption{Detection rate and false positive rate of the intrusion detection systems used in the framework}
\label{tab:6}
\resizebox{\linewidth}{!}{%
\begin{tabular}{>{\hspace{0pt}}p{0.301\linewidth}>{\hspace{0pt}}p{0.297\linewidth}>{\hspace{0pt}}p{0.397\linewidth}}
\hline
\textbf{~ ~IDS} & \textbf{~ DR (\%)} & \textbf{~ ~ ~FPR (\%)}  \\
\hline
Anomaly         & ~ ~95.01           & ~ ~ ~ ~~9.97            \\
\hline
~Misuse         & ~ ~50.04           & ~ ~ ~ ~ 0.9             \\
\hline
~Hybrid         & ~ ~97.2            & ~ ~ ~ ~ 0.88            \\
\hline
\end{tabular}
}
\end{table}
\section{Conclusion and future works}
\label{sec:6}
Our goal in this paper is to design a healthcare system based on wireless sensor networks and secure it against unauthorized access and network attacks. Consequently, we create a healthcare system with new capabilities. In this system, we consider security measures appropriate to wireless sensor networks besides providing medical services. For this purpose, after defining the agents and their interactions, the intrusion detection systems are used to confront healthcare network attacks. These intrusion detection systems are proportional to the level of energy and the sensitivity of data of each layer in our framework. Also, the implementation and evaluation of each IDS of the layers prove the correctness of our framework design. We also discuss SVM as an efficient network traffic classifier in our framework.
Our proposed framework consists of two steps, so we can implement each step with other methods and tools and compare the results. Consequently, we suggest the following items as the future works: 1) Using a standard medical dataset in the multiagent design phase. 2) Using other learning algorithms to set the rules of the patient agent and the ambient agent to learn the usual behavior of patients and ambient sensors. It might enhance the flexibility of the agents.
%

%
\end{document}